\documentclass[amsmath,amssymb,12pt,superscriptaddress]{revtex4}
\usepackage{graphicx}
\usepackage{dcolumn}
\usepackage{bm}
	
\newcommand{\bolm}[1]{\mbox{\boldmath{$#1$}}}

\begin{document}

\thispagestyle{empty}

{\baselineskip0pt
\leftline{\large\baselineskip16pt\sl\vbox to0pt{
               \hbox{\it Department of Mathematics and Physics}
               \hbox{\it Osaka City  University}
               \vspace{1mm}
               \hbox{\it Yukawa Institute for Theoretical Physics} 
               \hbox{\it Kyoto University}\vss}}
\rightline{\large\baselineskip16pt\rm\vbox to20pt{\hbox{OCU-PHYS-272}
            \hbox{AP-GR-46}
            \hbox{YITP-08-09}
\vss}}%
}
\vskip3cm

\title{Horizons of Coalescing Black Holes on Eguchi-Hanson Space}

\author{Chul-Moon Yoo\footnote{E-mail:yoo@yukawa.kyoto-u.ac.jp}}
\affiliation{
Yukawa Institute for Theoretical Physics, Kyoto University, 
Kyoto 606-8502, Japan
}
\affiliation{ 
Department of Mathematics and Physics,
Graduate School of Science, Osaka City University,
3-3-138 Sugimoto, Sumiyoshi, Osaka 558-8585, Japan
}
\author{Hideki Ishihara\footnote{E-mail:ishihara@sci.osaka-cu.ac.jp}}
\author{\\ Masashi Kimura\footnote{E-mail:mkimura@sci.osaka-cu.ac.jp}}
\author{Ken Matsuno\footnote{E-mail:matsuno@sci.osaka-cu.ac.jp}}
\author{Shinya Tomizawa\footnote{E-mail:tomizawa@sci.osaka-cu.ac.jp}}
\affiliation{ 
Department of Mathematics and Physics,
Graduate School of Science, Osaka City University,
3-3-138 Sugimoto, Sumiyoshi, Osaka 558-8585, Japan
}
\date{April 9, 2008}

\begin{abstract}
Using the numerical method, 
we study dynamics of 
coalescing black holes on the Eguchi-Hanson base space.
Effects of a difference in spacetime topology on the 
black hole dynamics is discussed. 
We analyze appearance and disappearance process of marginal surfaces. 
In our calculation, 
the area of a coverall 
black hole horizon at the creation time 
in the coalescing black holes solutions 
on Eguchi-Hanson space is larger than that in the 
five-dimensional Kastor-Traschen solutions. 
This fact suggests that 
the black hole production on the Eguchi-Hanson space 
is easier than that on the flat space. 
\end{abstract}


\maketitle

\pagebreak

\section{Introduction}

In the framework of the brane world scenario, 
higher dimensional black holes are expected to be produced 
in a future linear collider
~\cite{Banks:1999gd,Dimopoulos:2001en,Giddings:2001bu,IdaOdaPark1,IdaOdaPark2,IdaOdaPark3}.
By observing physical phenomena associated with the black holes 
we might obtain 
evidences for existence of extra-dimensions. 
Such black holes, 
which evaporate by the Hawking radiation, 
are also expected to play crucial roles 
in the yet unaccomplished theoretical development to reconcile 
gravitational interactions with quantum description of nature. 

So far, many authors have focused mainly on asymptotically flat 
and stationary higher dimensional black holes since
they would be idealized models if such black holes 
are small enough for us to neglect the tension of a brane
or the size of extra dimensions. 
It has been clarified that such asymptotically flat higher dimensional black hole solutions have 
richer structure than the four-dimensional one
~\cite{Cai:2001su,MyersPerry,Emparan:2001wn,Galloway:2005mf}. 
However, 
there is no 
reason to restrict the asymptotic structures of 
higher dimensional spacetimes to the flat spacetime.
Then, we do not have to restrict ourselves to black hole solutions with asymptotic flatness. In fact, higher dimensional black holes would admit a variety of 
asymptotic structures.
For example, the black hole solutions in Kaluza-Klein theory
admit the structure of 
a twisted ${\rm S}^1$ fiber bundle over four-dimensional Minkowski spacetime~\cite{DM,GW,IM}
or
a direct product of ${\rm S}^1$ and four-dimensional Minkowski spacetime~\cite{MyersKKBH}.

Recently, 
the coalescing black holes solutions 
on Eguchi-Hanson space (CBEH) are constructed 
in the five-dimensional Einstein-Maxwell theory 
with a positive cosmological constant~\cite{Ishihara:2006ig}. 
These solutions are  asymptotic locally de Sitter spacetime; 
the topology of the radial coordinate $r=$const. surfaces is 
not a sphere $\rm S^3$ but the lens space. 
In this article, 
the behaviour of black holes  at the early time and the late time 
are mainly discussed. 
The reason for this restriction is that 
it is easy to analyze the structure of solutions in such the region, which
one can regard 
as that of the five-dimensional Reissner-Nordstr\"om-de Sitter solution (RNdS). 
As a result, it is clarified that the solutions describe the physical situation such that 
two black holes with the topology of $\rm S^3$ 
coalesce and change into a single black hole with 
the topology of the lens space $L(2;1)={\rm S}^3/{\mathbb Z}_2$.

Another solution of Einstein-Maxwell theory
with a positive cosmological constant in arbitrary dimensions 
had been already found by London~\cite{London:1995ib}. 
These solutions, which are the generalization of 
the Kastor-Traschen solution~\cite{Kastor:1992nn} 
to higher dimensions, describe the dynamical situation 
such that the arbitrary number of multi-black holes 
with a spherical topology coalesce into a single black hole with a spherical topology 
in asymptotically de Sitter spacetime. 
Two black holes case of 
the five-dimensional Kastor-Traschen solutions (5DKT) describes that 
the two black holes with ${\rm S}^3$ coalesce into a single black hole with ${\rm S}^3$.

The purpose of this article is to investigate the global structure of 
the CBEH and the 5DKT
by the numerical approach
and to clarify the effects on coalescence of black holes
brought about by the difference in asymptotic structure between both solutions.
Following the numerical method in Refs.
\cite{Nakao:1994mm,Sasaki:1980,1974AnPhy..83..449C}, 
where they discussed how marginal surfaces evolve with time 
in the four-dimensional Kastor-Traschen solutions, 
we numerically investigate the existence and the time evolution of 
marginal surfaces. Especially, 
we focus on the appearance and disappearance process of marginal surfaces.
We also discuss the time evolution of these areas.

The rest of this article is organized as follows. 
In Sec. \ref{sec:2}, we review 
the CBEH and the 5DKT.
We show the method to search for marginal surfaces in Sec. \ref{sec:3}. 
Then, time sequence of marginal surfaces and these areas are shown in Sec. \ref{sec:4}. 
Sec. \ref{sec:5} is devoted to the summary and discussion.

\section{Brief Review}\label{sec:2}
\subsection{five-dimensional Kastor-Traschen solutions}
First, let us consider the 5DKT~\cite{London:1995ib}, namely, the black hole solutions on a flat base space.
Especially, we concentrate on the solution with two black holes whose masses are $m_1$ and $m_2$ at early time
\begin{equation}
ds^2=a^2\left[-H_{\rm KT}^{-2}d\tau^2+H_{\rm KT}\left(dr^2+r^2d\Omega^2_{\rm S^3}
\right)\right], \label{eq:ktmet}
\end{equation}
where $H_{\rm KT}$ is given by 
\begin{equation}
H_{\rm KT}=\lambda\tau+\frac{m_1}{|\bolm r-\bolm r_1|^2}
+\frac{m_2}{|\bolm r-\bolm r_2|^2}, 
\end{equation}
with the position vector on the four-dimensional Euclid space 
$\bolm r $. 
$\bm r_1$ and $\bm r_2$ are the positions of point sources. 
We can set $\bolm r_1 =(0,0,0,1)$ and $\bolm r_2 =(0,0,0,-1)$ 
without a loss of 
generality.

\subsubsection{Early time}
Let us focus on the neighbourhood of $\bolm r=\bolm r_i~(i=1,2)$. 
In the new coordinate  $\tilde{ r}=| \bolm r-\bolm r_i|$, 
we can write the metric (\ref{eq:ktmet}) as 
\begin{eqnarray}
ds^2
&\simeq &a^2\left[
 -\biggl(\lambda \tau+\frac{m_i}{\tilde r^2}\biggr)^{-2}d\tau^2
+\biggl(\lambda \tau+\frac{m_i}{\tilde r^2}\biggr)
\biggl\{d\tilde r^2+\tilde r^2d\Omega_{\rm S^3}^2\biggr\}\right],\label{eq:near}
\end{eqnarray}
where $d\Omega_{\rm S^3}^2$ is the metric of a unit three-sphere.
This is identical to the metric of 
the RNdS 
with mass parameter $m_i$ 
except for the conformal factor $a^2$ which 
does not contribute to the horizon condition $\theta_{\rm out}=0$, 
where $\theta_{\rm out}$ is the out-going null expansion on 
the $\tau=$const and $\tilde r=$const surface. 

For this metric, 
let us introduce a variable $ x :=\lambda \tau \tilde r^2$, 
and then horizons occur at $ x$ satisfying 
\begin{eqnarray}
\lambda ^2( x+m_i)^3-4 x^2=0.\label{eq:cubic}  
\end{eqnarray}
For $m_i < 
16/(27\lambda ^2)$, 
there are three horizons, i.e., 
the inner and outer black hole horizons 
and the de Sitter horizon,
which correspond to the three real roots 
$ x_{\rm in}[m_i]< x_{\rm BH}[m_i]< x_{\rm dS}[m_i]$, respectively.

If $m_i<16/(27\lambda ^2)~
(i=1,2)$, 
the horizon radius 
$\tilde r_{\rm BH}^2:=x_{\rm BH}[m_i]/(\lambda \tau)$ satisfy 
$\tilde r_{\rm BH}\ll|\bolm r_i|=1$ at an early time $\tau \ll 0$. 
This fact means that we can find an approximately spherical and sufficiently small 
black hole horizon around $\bolm r=\bolm r_i$.
Hence, an outer trapped region always exists around ${\bm r}={\bm r}_i$

\subsubsection{Late time}
Next, we study the asymptotic behaviour of the metric for large $r:=|{\bm r}|$,  where we assume that $r$ is much larger than the coordinate distance $|{\bm r_1}-{\bm r_2}|=2$ between the two masses $m_1$ and $m_2$.
Then, the metric takes the following form,
\begin{eqnarray}
ds^2
&\simeq &a^2\left[
 -\biggl(\lambda \tau+\frac{m_1+m_2}{  r^2}\biggr)^{-2}d\tau^2
+\biggl(\lambda \tau+\frac{m_1+m_2}{  r^2}\biggr)
\biggl\{d  r^2+  r^2d\Omega_{\rm S^3}^2\biggr\}\right].\label{eq:far}
\end{eqnarray}
This metric resembles 
that of the RNdS 
with mass equal to $m_1+m_2$. 
If we assume $m_1+m_2<16/(27\lambda ^2)$, 
the horizon radius 
$r_{\rm BH}^2:=x_{\rm BH}[m_1+m_2]/(\lambda \tau)$ satisfy 
$r_{\rm BH}\gg |\bolm r_1-\bolm r_2|=2$ at late time $\tau\to -0$. 
Then the approximate form of the metric (\ref{eq:far}) is 
valid around $r=r_{\rm BH}$. 
Hence, 
an approximately spherical 
black hole horizon can be found around $r=r_{\rm BH}$.
in the metric (\ref{eq:ktmet}).

\subsection{Black holes on Eguchi-Hanson base space}

Second, we give the brief review on the CBEH~\cite{Ishihara:2006ig} 
whose metric is given by 
\begin{equation}
ds^2=a^2\left[-H_{\rm EH}^{-2}d\tau^2+\frac{1}{8}H_{\rm EH}
\left\{V^{-1}dR^2+V^{-1}R^2
d\Omega_{\rm S^2}^2
+V
\left(d\psi+\omega_\phi d\phi\right)^2\right\}\right], \label{eq:ehmet}
\end{equation}
where 
\begin{eqnarray}
H_{\rm EH}&=&\lambda\tau+\frac{2m_1}{\left|\bolm R-\bolm R_1\right|}
+\frac{2m_2}{\left|\bolm R-\bolm R_2\right|},\\
V^{-1}&=&\frac{1}{\left|\bolm R-\bolm R_1\right|}
+\frac{1}{\left|\bolm R-\bolm R_2\right|}, \\
\omega_\phi&=&\frac{z-1}{\left|\bolm R-\bolm R_1\right|}
+\frac{z+1}{\left|\bolm R-\bolm R_2\right|},  \\
d\Omega_{\rm S^2}^2
&=& d\theta^2+\sin^2\theta d\phi^2, 
\end{eqnarray}
and $\bolm R=(R\sin\theta\cos\phi,R\sin\theta\sin\phi,R\cos\theta)$ 
is the position vector on the three-dimensional 
Euclid space and positions of point sources $\bolm R_1$ and $\bolm R_2$ 
are set to be $(0,0,1)$ and $(0,0,-1)$. 
The range of angular coordinates is defined by 
$0\leq\theta\leq\pi$, $0\leq\phi \leq 2\pi$ and $0\leq\psi \leq 4\pi$. 
This metric is given by equation (9) in Ref.\cite{Ishihara:2006ig}, 
rewriting as $R\rightarrow aR$, $\tau\rightarrow a\tau$, 
$\lambda\rightarrow \lambda/a$, $m_1\rightarrow a^2m_1$ 
and $m_2\rightarrow a^2m_2$. 
This is a solution of 
the five-dimensional Einstein equation with a positive cosmological constant 
and the Maxwell equation with a gauge potential one-form 
given by 
\begin{eqnarray}
\quad \bm{A}
=\pm\frac{\sqrt{3}}{2}aH_{\rm EH}^{-1}d\tau. 
\end{eqnarray}
In order to focus on the coalescence of two black holes, we consider only the contracting phase $\lambda<0$. Though $\tau$ runs the range $(-\infty,\infty)$, in this article we investigate only the region $-\infty< \tau \le 0$.

\subsubsection{Early time}
 First,
let us focus on the neighbourhood of ${\bm R}={\bm R_i}\ (i=1,2)$.  
In terms of the new coordinate $\bar r^2:=|\bolm{R}-\bolm{R}_i|/2$, 
the metric can be written in the form,
\begin{equation}
ds^2\simeq a^2\left[-\left(\lambda \tau+\frac{m_i}{\bar r^2}\right)^{-2}
d\tau^2+\left(\lambda\tau+\frac{m_i}{\bar r^2}\right)
\left\{d\bar r^2+\frac{\bar r^2}{4}d\Omega_{{\rm S}^2}^2+\frac{\bar r^2}{4}
\left(d\psi+\cos\theta d\phi\right)^2\right\}\right]. \label{eq:neareh}
\end{equation}
This is equivalent to the metric of 
the the RNdS 
which has the mass equal to $m_i$ written in the cosmological coordinate. 
Hence 
like the 5DKT, 
we can conclude that 
a nearly spherical and small 
black hole horizon can be found
around $\bolm R=\bolm R_i$ in the 
metric (\ref{eq:ktmet}) 
at the early time, and sufficiently small spheres with the topology 
of $S^3$ centered at ${\bm R}={\bm R}_i$ are always outer trapped.

\subsubsection{Late time}
Next, we study the asymptotic behaviour of the metric (\ref{eq:ehmet}) 
in the region where 
$R$ is much larger than 
the coordinate distance $|{\bm R_1}-{\bm R_2}|=2$. 
Here, let us introduce a new coordinate $\hat r^2:=R$, 
and then the metric takes the following form
\begin{equation}
ds^2\simeq a^2\left[-\left(\lambda \tau+\frac{m}{\hat r^2}\right)^{-2}
d\tau^2+\left(\lambda\tau+\frac{m}{\hat r^2}\right)
\left\{d\hat r^2+\frac{\hat r^2}{4}d\Omega_{\rm S^2}^2+\frac{\hat r^2}{4}
\left(\frac{1}{2}d\psi+\cos\theta d\phi\right)^2\right\}\right], 
\label{eq:fareh0}
\end{equation}
where $m=2(m_1+m_2)$. 
This resembles the metric of the RNdS solution with mass equal to $m$, 
and 
if we assume $m<16/(27\lambda ^2)$, 
a nearly spherical 
black hole horizon can be found in the metric (\ref{eq:ehmet}) with $\hat r=r_{\rm BH}[m]$  
at late time $\tau\to -0$. 

However, we note that the metric form of (\ref{eq:fareh0}) differ from that of 
the RNdS solution 
in the following point; 
Each $\hat r = {\rm const}$ surface is topologically the lens space 
$L(2;1)={\rm S}^3/{\mathbb Z}_2$, 
while it is diffeomorphic to ${\rm S}^3$ in the RNdS solution.
We can regard ${\rm S}^3$ and the lens space $L(2;1)={\rm S}^3/{\mathbb Z}_2$ 
as examples of Hopf bundles, i.e., 
${\rm S}^1$ bundle over ${\rm S}^2$. 
The difference between these metrics appears in Eqs.(\ref{eq:neareh}) 
and (\ref{eq:fareh0}): 
$d\psi$ in the 
 metric (\ref{eq:neareh}) 
is replaced by $d\psi/2$ in the 
 metric (\ref{eq:fareh0}). 
Therefore, at late time, the topology of the trapped surface 
is the lens space $L(2;1)={\rm S}^3/{\mathbb Z}_2$
in the metric (\ref{eq:ehmet}).

\subsection{Comparison}
The above results suggest that 
both solutions describe the coalescence of black holes. 
(In fact, using the numerical techniques, Nakao et.al. showed that 
the four-dimensional Kastor-Traschen solutions describe
such physical process~\cite{Nakao:1994mm}.)
Between both solutions, 
there exists the essential difference, namely, 
in the 5DKT, 
two black holes with the topology of ${\rm S}^3$ coalesce 
into a single black hole with the topology of ${\rm S}^3$, while in the CBEH,
two black holes with the topology of ${\rm S}^3$ 
coalesce and change into a single black hole with the topology 
of $L(2;1)={\rm S}^3/{\mathbb Z}_2$. 
In the next section, we investigate how two black holes coalesce 
in the two solutions by pursuing time evolution of marginal surfaces.

\section{Method to search for marginal surfaces}\label{sec:3}

Here, we seek marginal surfaces on $\tau={\rm constant}$ surfaces, 
which are defined as surfaces of co-dimension two such that 
the out-going orthogonal null geodesics have zero convergence 
$\theta_{\rm out}$ 
on the surfaces. 
The metrics (\ref{eq:ktmet}) and (\ref{eq:ehmet2}) are decomposed into the form
\begin{eqnarray}
g_{ab}=-n_an_b+h_{ab},
\end{eqnarray}
where $n^a:=H_{\rm EH,KT} a^{-1}(\partial/\partial \tau)^a$
and $h_{ab}$ denote the timelike unit vector normal to the
$\tau={\rm constant}$ surfaces and the induced metric on the surfaces, 
respectively.

In our numerical computation, 
we use the coordinate system $(\tau,z,\rho,\phi,\psi)$. 
The metric (\ref{eq:ktmet}) is written as 
\begin{equation}
ds^2=a^2\left[-H_{\rm KT}^{-2}d\tau^2+H_{\rm KT}\left\{dz^2+d\rho^2
+\rho^2\left(d\phi^2
+\sin^2\phi d\psi^2\right)\right\}\right], \label{eq:ktmet2}
\end{equation}
in this coordinate system, and 
the metric (\ref{eq:ehmet}) is written as 
\begin{equation}
ds^2=a^2\left[-H_{\rm EH}^{-2}d\tau^2+\frac{1}{8}H_{\rm EH}
\left\{V^{-1}\left(dz^2+
d\rho^2+\rho^2 d\phi^2\right)+V
\left(d\psi+\omega_\phi d\phi\right)^2\right\}\right].  \label{eq:ehmet2}
\end{equation}

Let $s^a$ be the spacelike unit vector normal to such 
marginal surfaces on the $\tau=$constant surfaces, 
and consider the marginal surfaces as  
$(z,\rho,\phi,\psi)=\left(z(v),\rho(v),\phi,\psi\right)$, 
i.e., they are parameterized by $v,\phi$ and $\psi$ on the $\tau=$constant surfaces. 
Then, the metric $h_{ab}$ can be written on the following form
\begin{eqnarray}
h_{ab}=s_as_b+\delta_{ij}(e^i)_{a} (e^j)_b, 
\end{eqnarray}
where $\delta_{ij}={\rm diag}(1,1,1)$ and $(e_i)^a\ (i=1,2,3)$ are 
triplet bases on the marginal surface. 
In the case of the CBEH, we can set $s^a$ and $(e_i)^a$
in the forms 
\begin{eqnarray} 
&&(e_{1})^a=\frac{2\sqrt{2V}}{a\sqrt{H_{\rm EH}
(\dot \rho^2+\dot z^2)}}\left(\dot z\left(\frac{\partial }
{\partial z}\right)^a
+\dot \rho\left(\frac{\partial }
{\partial \rho}\right)^a \right),\label{eq18}\\
&&
(e_{2})^a
=
\frac{2\sqrt{2V}}{a\sqrt{H_{\rm EH}(\rho^2+ \omega_{\phi}^2 V^2)}}
\left(\frac{\partial }{\partial \phi}\right)^a,
\\
&&(e_3)^a=\frac{2\sqrt{2}}{a\rho\sqrt{H_{\rm EH}V}}
\left(\frac{
-\omega_{\phi}V^2}{\sqrt{\rho^2+\omega_{\phi}^2V^2}}
\left(\frac{\partial }{\partial \phi}\right)^a
+\sqrt{\rho^2+\omega_{\phi}^2V^2}
\left(\frac{\partial}{\partial \psi}\right)^a\right), \\
&& s^a=\pm\frac{2\sqrt{2V}}{a\sqrt{H_{\rm EH}
(\dot \rho^2+\dot z^2)}}
\left(\dot \rho\left(\frac{\partial }{\partial z}\right)^a
-\dot z\left(\frac{\partial }{\partial \rho}\right)^a\right),
\label{eq21}
\end{eqnarray}
where the sign $\pm$ of $s^a$ should be chosen so 
that $s^a$ directs outward.
On the other hand, in the case of the 5DKT, 
we can set these vectors in the forms 
\begin{eqnarray}
&&(e_1)^a=\frac{1}{a\sqrt{H_{\rm KT}(\dot \rho^2+\dot z^2)}}
\left(\dot z \left(\frac{\partial }{\partial z}\right)^a
+\dot \rho \left(\frac{\partial }{\partial \rho}\right)^a\right), 
\label{eq22}\\
&&(e_2)^a=\frac{1}{a\rho\sqrt{H_{\rm KT}}}
\left(\frac{\partial }{\partial \phi}\right)^a, \\
&&(e_3)^a=\frac{1}{a\rho\sqrt{H_{\rm KT}}\sin\phi}
\left(\frac{\partial }{\partial \psi}\right)^a, \\
&&s^a=\pm\frac{1}{a\sqrt{H_{\rm KT}(\dot \rho^2+\dot z^2)}}
\left(\dot \rho \left(\frac{\partial }{\partial z}\right)^a
-\dot z\left(\frac{\partial }{\partial \rho}\right)^a\right).
\label{eq25}
\end{eqnarray}

The expansion $\theta_{\rm out}$ of 
the null congruence which is normal to the marginal surface 
is given by 
\begin{equation}
\theta_{\rm out}
=(h^{ab}-s^a s^b)\nabla_b(n_a+s_{ a})
=-\sum_{i=1,2,3}s_a(e_i)^bD_b(e_i)^a+s^as^bk_{ab}-{\rm tr}k, 
\label{eq:nullex}
\end{equation}
where $k$ is the trace of the extrinsic curvature $k_{ab}$ of 
the $\tau=$constant surface. 
By the definition of a marginal surface, the expansion vanishes 
$\theta_{\rm out}=0$ on the surface. 
On the other hand, the expansion of the ingoing null congruence which is normal to 
the marginal surface is given by 
\begin{equation}
\theta_{\rm in}
=(h^{ab}-s^a s^b)\nabla_b(n_a-s_{ a})
=\sum_{i=1,2,3}s_a(e_i)^bD_b(e_i)^a+s^as^bk_{ab}-{\rm tr}k. 
\label{eq:innullex}
\end{equation}
Since $\theta_{\rm out}$ vanishes on the marginal surface by its definition, 
we have 
\begin{equation}
\theta_{\rm in}
=\theta_{\rm in}+\theta_{\rm out}
=2s^as^bk_{ab}-2{\rm tr}k=\frac{3\lambda}{a}<0. 
\label{eq:innullex2}
\end{equation}

Equations \eqref{eq18}-\eqref{eq21} or Eqs.\eqref{eq22}-\eqref{eq25} 
with Eq.\eqref{eq:nullex} gives a second-order ordinary differential 
equation 
\begin{equation}
\theta_{\rm out}\left(\ddot z,\ddot \rho,\dot z,\dot \rho\right)=0
\label{diffeq}
\end{equation}
for marginal surfaces on $\rho-z$ plane. 
We find smooth closed curves on the $\rho-z$ plane, 
$\rho=\rho(v),~ z=z(v)$, satisfying Eq.\eqref{diffeq}. 
It should be noted that Eq.\eqref{diffeq} does not depend on 
the parameter $a$. 
By use of the freedom in the choice
of the parameter $v$, following Cadez\cite{Nakao:1994mm,1974AnPhy..83..449C}, 
we fix $v$ by 
\begin{equation}
\dot z^2+\dot \rho^2=\left(\frac{8V}{H_{\rm EH}}\right)^2, \label{eq:deq2}
\end{equation}
in the case of the CBEH, and 
\begin{equation}
\dot z^2+\dot \rho^2=H_{\rm KT}^{-2}. \label{eq:deq3}
\end{equation}
in the case of the 5DKT.

Using these parametrization (\ref{eq:deq2}) and (\ref{eq:deq3}) of 
$v$ and imposing the equations on the
boundary conditions $\dot z=0$ at $z$-axis,  
we can numerically search for marginal surfaces.


\section{Time evolution of horizons}\label{sec:4}
\subsection{Marginal Surfaces}

We would like to pursue how two 
black holes evolve with time and coalesce for both solutions. 
We restrict the range of the mass parameters to
\begin{equation}
m_1+m_2<\frac{8}{27\lambda^2} \label{eq:ineq}
\end{equation}
so that a black hole horizon exists after the coalescence. 
In this article, we consider the case where 
two black holes at early time have equal masses and 
we set them to be $m_1=m_2=1/(8\lambda^2)$ and 
$\lambda=-1/(2\sqrt{2})$ for both solutions. 
Under this assumption, since there is a reflection symmetry $z \to -z$, 
it is sufficient to consider only the region of $z\ge0$.
In general, several marginal surfaces exist on each time slice. 
We label each marginal surface which corresponds to a 
black hole horizon and de Sitter horizon 
at the early time 
as ${\rm BH_E}$ and ${\rm dS}_{\rm E}$, respectively, 
and label each marginal surface 
which corresponds to the black hole horizon and de Sitter horizon 
at the late time as ${\rm BH}_{\rm L}$ and ${\rm dS}_{\rm L}$, respectively. 
Some of marginal surfaces appear or disappear 
in pairs with another marginal surface. 
We label the marginal surfaces other than black hole horizons and 
de Sitter horizons as ${\rm MS}_i\ (i=1,2,\cdots)$.  
To avoid confusion, we do not depict marginal surfaces 
which are not related to the appearance and disappearance of ${\rm BH_E}$,
${\rm dS_E}$, ${\rm BH_L}$ and ${\rm dS_L}$.

\begin{figure}[htbp]
\begin{center}
\includegraphics[scale=1]{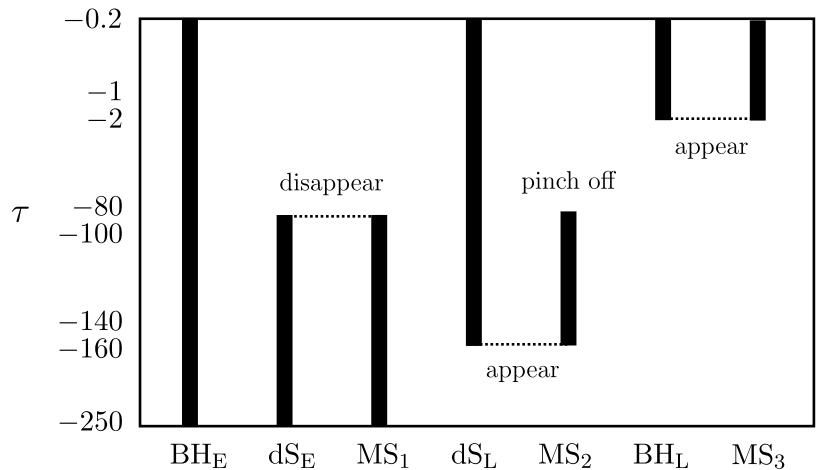}
\caption{Time evolution of marginal surfaces in the 5DKT: 
This figure shows when $\rm BH_E$, $\rm dS_E$, $\rm BH_L$, $\rm dS_L$ and three marginal surfaces $\rm MS_1$, $\rm MS_2$, $\rm MS_3$ exist. 
The vertical axis denotes the values of $\tau$.
A pair of marginal surfaces connected by a dashed line
appears or disappears at one time.
}
\label{fig:tabkt}
\end{center}
\end{figure}

\begin{figure}[htbp]
\begin{center}
\includegraphics[scale=0.87]{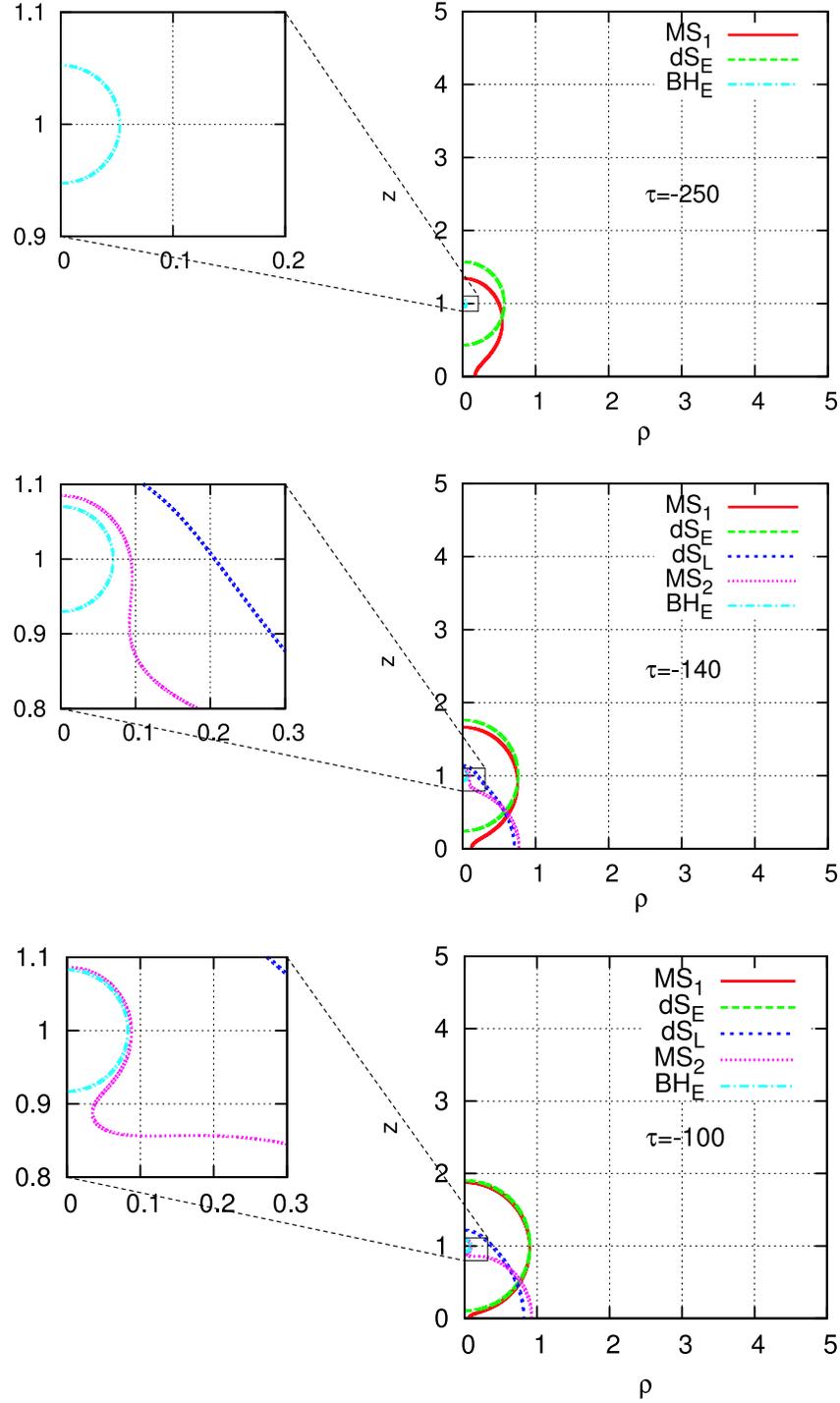}
\caption{
Time evolution of marginal surfaces in the 5DKT in $-250 < \tau < -100$. 
The each left frame is the scaled-up figure of the region near ${\rm BH_E}$.
At $\tau = -250$, ${\rm BH_{E}}$, ${\rm dS_{E}}$ and $\rm MS_1$ exist.
${\rm dS_L}$ and $\rm{MS_2}$ appear at same time during $-160 < \tau < -140$.
}
\label{fig:kthori1}
\end{center}
\end{figure}

\begin{figure}[htbp]
\begin{center}
\includegraphics[scale=0.87]{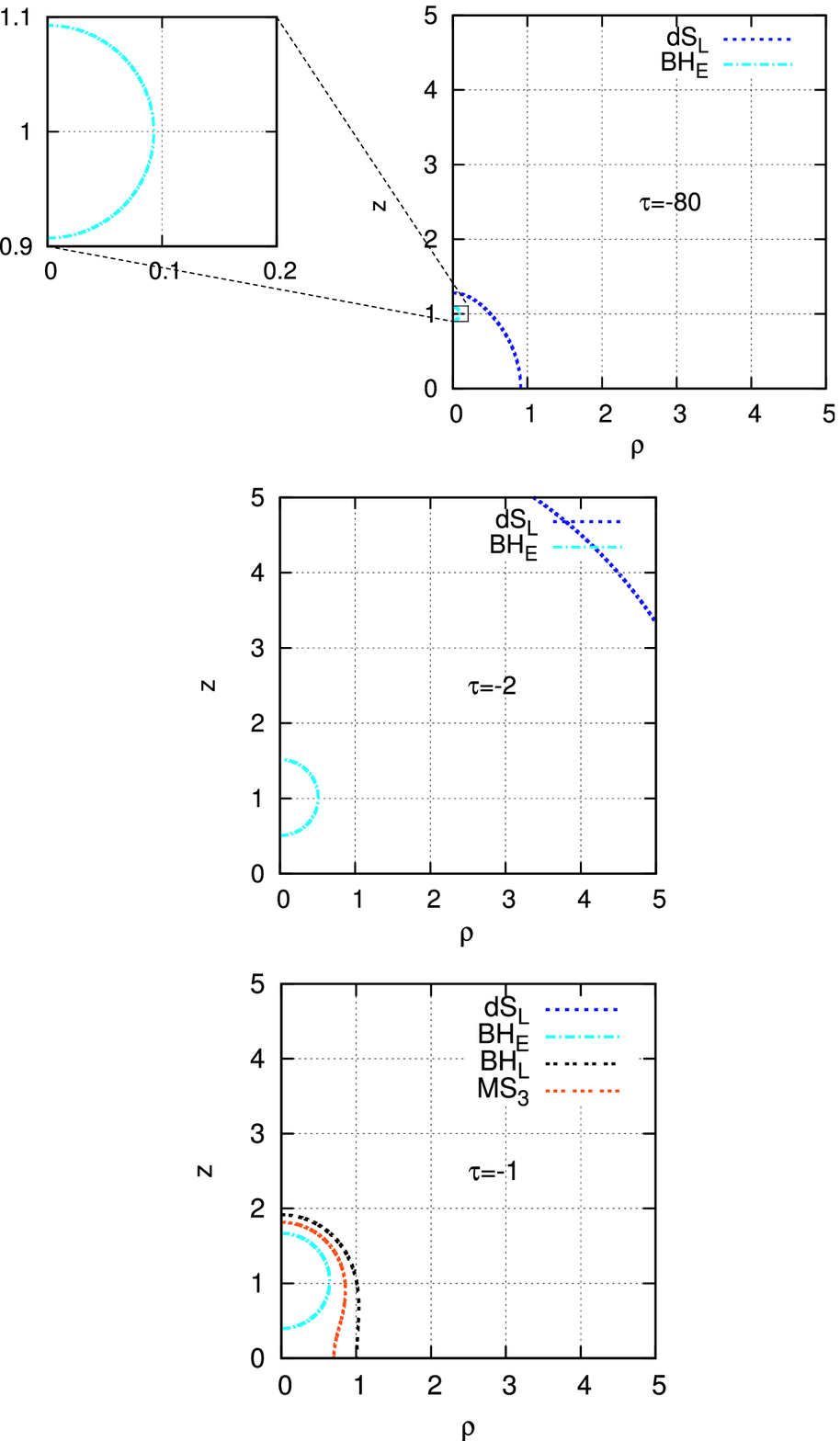}
\caption{
Time evolution of marginal surfaces in the 5DKT in $-80 < \tau < -1$. 
The left frame at $\tau = -80$ is the scaled-up figure of the region near ${\rm BH_E}$.
${\rm dS_E}$ and $\rm{MS_1}$ disappear at same time
and ${\rm MS_2}$ pinches off
during $-100 < \tau < -80$.
Finally ${\rm MS_3}$ and $\rm{BH_L}$ appear at same time during $-2 < \tau < -1$.
In the figure at $\tau=-1$, ${\rm dS_L}$ exists in 
the outside the frame. 
}
\label{fig:kthori2}
\end{center}
\end{figure}

FIG.\ref{fig:tabkt}, FIG.\ref{fig:kthori1} and FIG.\ref{fig:kthori2}
show the time sequence of marginal surfaces in the 5DKT.
Before $\tau=-250$, there are two black hole horizons ${\rm BH_E}$,
two de Sitter horizons ${\rm dS_E}$ 
enclosing each $\rm BH_E$. In addition, there is a marginal surface
${\rm MS_1}$ surrounding the two black hole horizons.
After the lapse of time, at a time within the period $-160< \tau<-140$,
another de Sitter horizon ${\rm dS_L}$ appears in pairs with another marginal surface ${\rm MS_2}$.
After a brief interval, each ${\rm dS_E}$ disappears in pairs with
${\rm MS_1}$, and ${\rm MS_2}$ pinches off at a time in $-100 < \tau < -80$.
Finally, at a time in
$-10<\tau<-5$, a new black hole horizon
${\rm BH_L}$ appears in pairs with a new marginal surface ${\rm MS_3}$,
and then it asymptotically approaches to the black hole horizon of the RNdS
with the mass parameter $m_1+m_2$.


\begin{figure}[htbp]
\begin{center}
\includegraphics[scale=1]{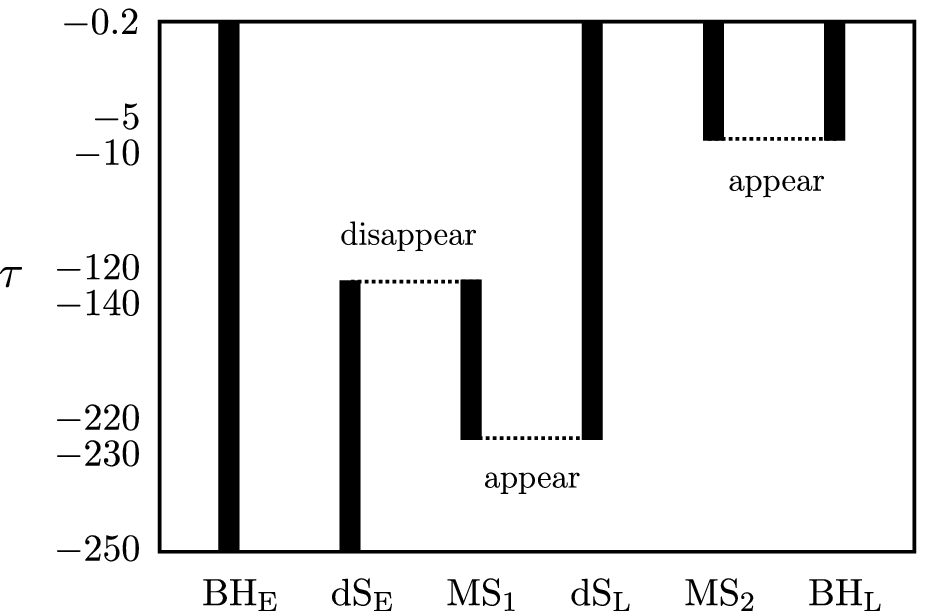}
\caption{Time evolution of marginal surfaces in the CBEH: This figure shows 
when $\rm BH_E$, $\rm dS_E$, $\rm BH_L$, $\rm dS_L$ and two marginal surfaces $\rm MS_1$, $\rm MS_2$ exist. 
The vertical axis denotes the values of $\tau$.
A pair of marginal surfaces connected by a dashed line
appears or disappears at one time.
}
\label{fig:tabeh}
\end{center}
\end{figure}

\begin{figure}[htbp]
\begin{center}
\includegraphics[scale=0.87]{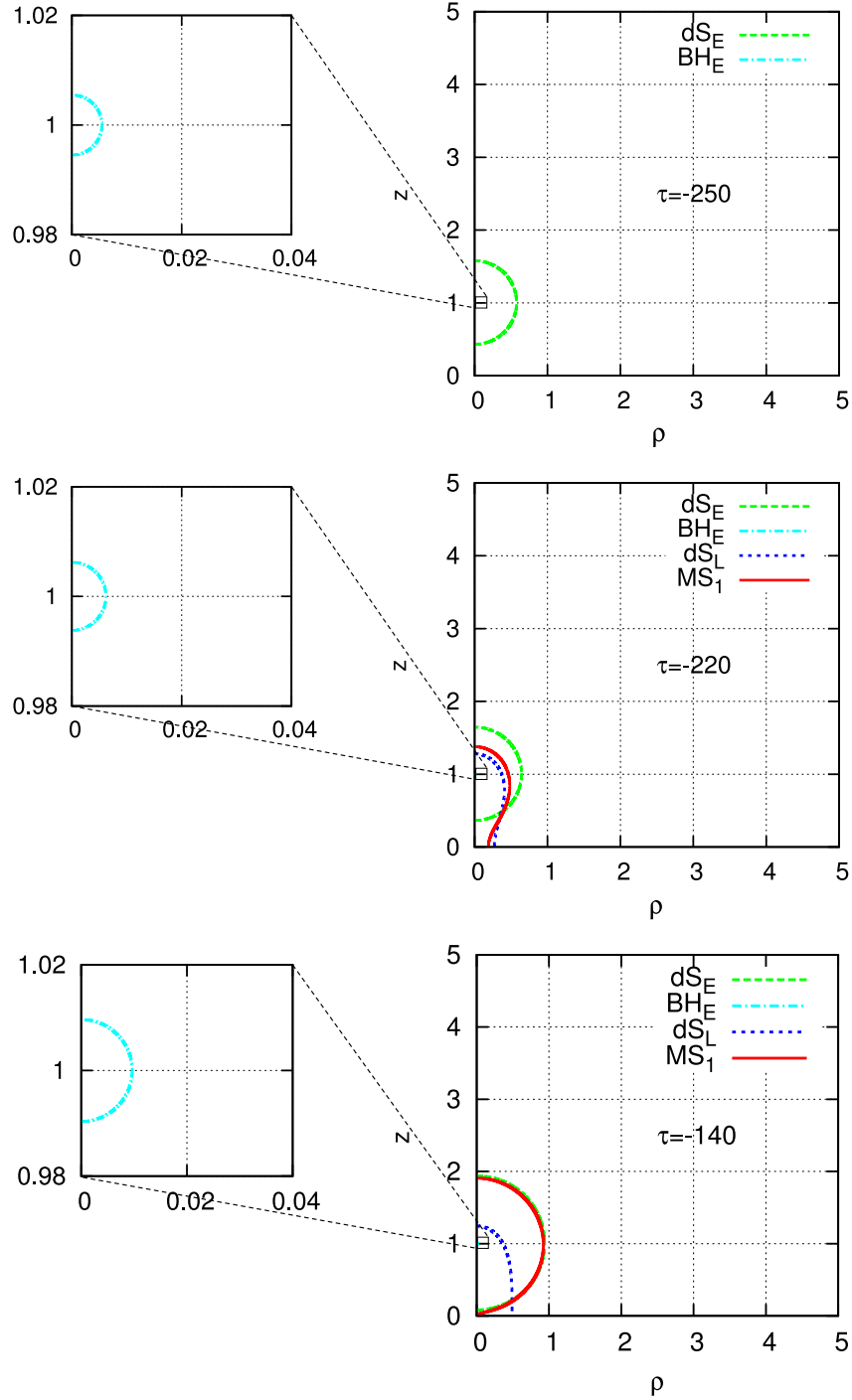}
\caption{Time evolution of marginal surfaces in the CBEH in $-250 < \tau < -140$. 
The each left frame is the scaled-up figure of the region near ${\rm BH_E}$.
At $\tau = -250$, ${\rm BH_{E}}$ and ${\rm dS_{E}}$ exist.
${\rm MS_1}$ and $\rm{dS_L}$ appear at same time during $-250 < \tau < -220$.
}
\label{fig:ehhori1}
\end{center}
\end{figure}

\begin{figure}[htbp]
\begin{center}
\includegraphics[scale=0.87]{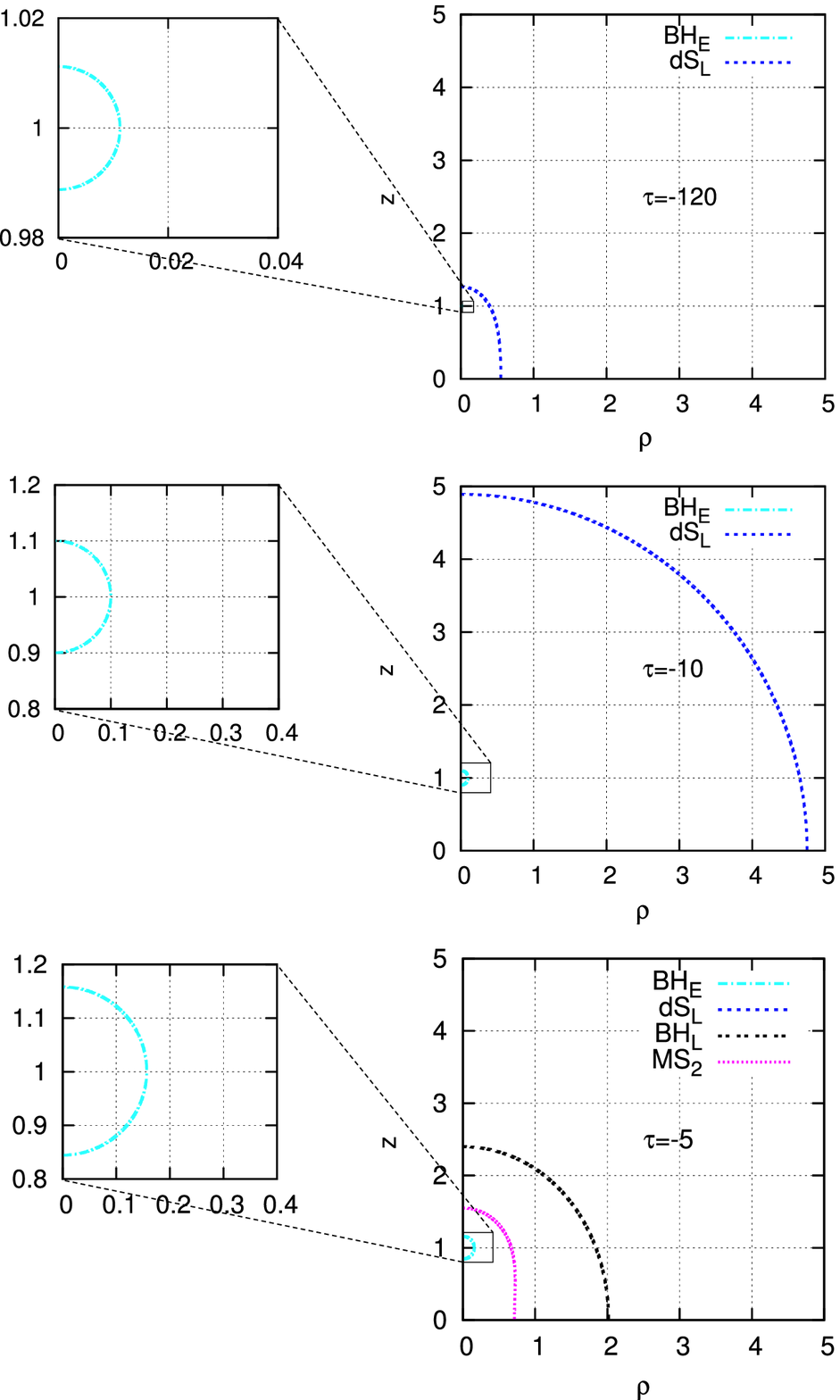}
\caption{Time evolution of marginal surfaces in the CBEH in $-120 < \tau < -5$. 
The each left frame is the scaled-up figure of the region near ${\rm BH_E}$.
${\rm dS_E}$ and $\rm{MS_1}$ disappear at same time during $-140 < \tau < -120$.
Finally ${\rm MS_2}$ and $\rm{BH_L}$ appear at same time during $-10 < \tau < -5$.
In the figure at $\tau=-5$, ${\rm dS_L}$ exists in 
the outside the frame. 
}
\label{fig:ehhori2}
\end{center}
\end{figure}

On the other hand, FIG.\ref{fig:tabeh}, FIG.\ref{fig:ehhori1} 
and FIG.\ref{fig:ehhori2} 
show the time sequence of marginal surfaces in the CBEH.
Before $\tau=-
250$, there exist two ${\rm BH_E}$ and two ${\rm dS_E}$.
At a time within the period 
$-230< \tau<-220$,
${\rm dS_L}$ appears in 
pairs with ${\rm MS_1}$.
After a brief interval, ${\rm dS_E}$ disappears in pairs with 
${\rm MS_1}$ at a time in $-140<\tau<-120$. 
Finally, ${\rm BH_L}$ appears in pairs with ${\rm MS_2}$ at a time in
$-10<\tau<-5$,
and ${\rm BH_L}$ approaches to a black hole horizon of 
the RNdS with the mass 
$2m_1+2m_2$ whose horizon topology is the lens space 
$L(2;1)={\rm S^3}/{\mathbb Z}_2$.


We can see that two solutions differ in the number of 
marginal surfaces which are related to appearance and 
disappearance of $\rm dS_E$ and $\rm dS_L$. 
In each solution, 
the situation does not essentially depend on 
the choice of the parameters $m_1$, $m_2$ 
and $\lambda$. 
Hence, this result suggests that this difference dose not come 
from the difference in the choice of the parameters 
but in the asymptotic structures.

\subsection{Areas of horizons}
%
First, for later convenience, we introduce 
the areas of horizons in
the RNdS with the horizon topology of $L(n;1)=\rm S^3/{\mathbb Z}_n$ given by 
\begin{eqnarray}
&&\mathcal{A}_n(r_{\rm BH}[m'])
=\frac{2\pi^2a^3r^3_{\rm BH}[m']}{n}
\left(\lambda\tau+\frac{m'}{r^2_{\rm BH}[m']}\right)^{3/2}, \\
&&\mathcal{A}_n(r_{\rm dS}[m'])
=\frac{2\pi^2a^3r^3_{\rm dS}[m']}{n}
\left(\lambda\tau+\frac{m'}{r^2_{\rm dS}[m']}\right)^{3/2}, 
\end{eqnarray}
where $m'$ is the mass parameter in the metric form written by 
\begin{equation}
ds^2= a^2\left[-\left(\lambda \tau+\frac{m'}{  r^2}\right)^{-2}
d\tau^2+\left(\lambda\tau+\frac{m'}{  r^2}\right)
\left\{d  r^2+\frac{  r^2}{4}d\Omega_{\rm S^2}^2+\frac{  r^2}{4}
\left(\frac{1}{n}d\psi+\cos\theta d\phi\right)^2\right\}\right]. 
\label{eq:fareh}
\end{equation}

In the previous work~\cite{Ishihara:2006ig}, 
we pointed out that after two black holes with the horizon topology of $\rm S^3$ coalesce, the area of the eventual single black hole 
in the CBEH is larger than that in the 5DKT, 
where we assume that each black hole in the CBEH
has the same mass and area as that in the 5DKT. 
The difference is essentially due to the asymptotic structure. 
While the 5DKT is asymptotically de Sitter and each surface enclosing two black holes has the topological structure of 
${\rm S}^3$, the topological structure of those in the CBEH is $L(2;1)={\rm S}^3/{\mathbb Z}_2$. 
In this sense, the CBEH is not asymptotically de Sitter but 
asymptotically locally de Sitter. 
Namely, the horizon radius of a black hole 
in the spacetime 
whose spatial infinity has the lens space $L(2;1)=\rm S^3/{\mathbb Z}_2$ 
becomes larger than that of the spacetime 
which has asymptotically Euclidean timeslices even if they 
have the same mass.

Using the results in the previous work, 
the ratios of areas of the black hole horizon 
and de Sitter horizon at the early time in the CBEH to 
those in the 5DKT become
\begin{eqnarray}
&&\frac{\mathcal{A}^{\rm EH}_{\rm BH}}
{\mathcal{A}^{\rm KT}_{\rm BH}}
=
\frac{\mathcal{A}_1(r_{\rm BH}[m_1])}{\mathcal{A}_1(r_{\rm BH}[m_1])}=1,\quad
\frac{\mathcal{A}^{\rm EH}_{\rm dS}}
{\mathcal{A}^{\rm KT}_{\rm dS}}
=
\frac{\mathcal{A}_1(r_{\rm dS}[m_1])}{\mathcal{A}_1(r_{\rm dS}[m_1])}=1, 
\label{eq:rateE}
\end{eqnarray}
where 
EH and KT denote the quantities associated 
with the CBEH 
and the 5DKT, respectively. 

On the other hand, those at the late time become 
\begin{eqnarray}
&&\frac{\mathcal{A}^{\rm EH}_{\rm BH}}{\mathcal{A}^{\rm KT}_{\rm BH}}
=
\frac{\mathcal{A}_2(r_{\rm BH}[2(m_1+m_2)])}
{\mathcal{A}_1(r_{\rm BH}[m_1+m_2])}
=C_1, 
\quad \frac{\mathcal{A}^{\rm EH}_{\rm dS}}{\mathcal{A}^{\rm KT}_{\rm dS}}
=
\frac{\mathcal{A}_2(r_{\rm dS}[2(m_1+m_2)])}
{\mathcal{A}_1(r_{\rm dS}[m_1+m_2])}=C_2, \label{eq:rateL}
\end{eqnarray}
where $C_1$ and $C_2$ are some constant determined by the values of 
$\lambda$, $m_1$ and $m_2$. 
In our setting, we find $C_1=0.332...$ and $C_2=2.350...$. 
Here, It should be noted that the area of the black hole horizon 
in the CBEH is larger than that 
in the 5DKT, but reversely the area of the de-Sitter horizon 
in the CBEH is smaller than that 
in the 5DKT.

Here, we numerically study how the areas of black hole horizons 
evolve with time in the 5DKT and the CBEH. 
The area of each marginal surface on 
$\tau$=constant surfaces is computed as
\begin{equation}
{\cal A}^{\rm EH}=\frac{\pi^2a^3}{2\sqrt{2}}\int^{v_1}_{v_2}\rho\sqrt{\frac{H_{\rm EH}^3(\dot z^2+\dot\rho^2)}{V}}dv, 
\quad {\cal A}^{\rm KT}=4\pi a^3\int^{v_1}_{v_2}\rho^2\sqrt{H_{\rm KT}^3(\dot z^2+\dot\rho^2)}dv. 
\end{equation}
where $v_1$ and $v_2$ satisfy $z(v_1)=z(v_2)=0$.

\begin{figure}[htbp]
\begin{center}
\includegraphics[scale=1.5]{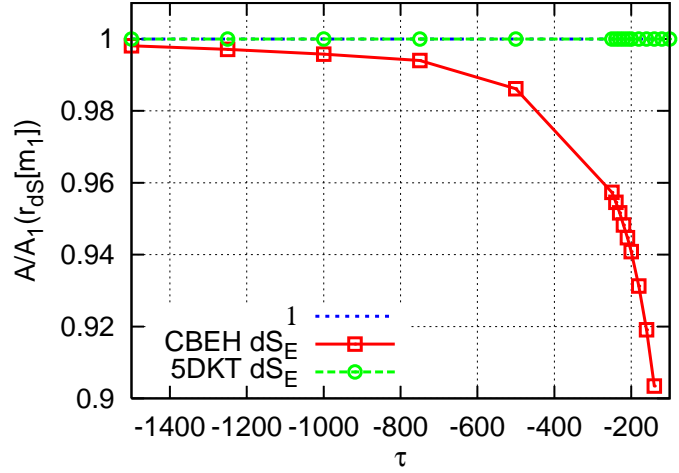}
\caption{Time evolution of the area
of ${\rm dS_E}$.
The vertical axis denotes the area normalized by ${\cal A}_1(r_{\rm dS}[m_1])$.
}
\label{fig:early-dS}
\end{center}
\end{figure}

\begin{figure}[htbp]
\begin{center}
\includegraphics[scale=1.5]{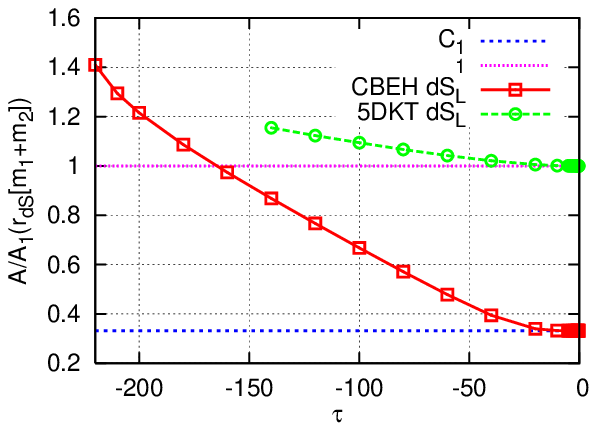}
\caption{Time evolution of the area
of ${\rm dS_L}$.
The vertical axis denotes the area normalized by ${\cal A}_1(r_{\rm dS}[m_1+m_2])$.
}
\label{fig:late-dS}
\end{center}
\end{figure}

\begin{figure}[htbp]
\begin{center}
\includegraphics[scale=1.5]{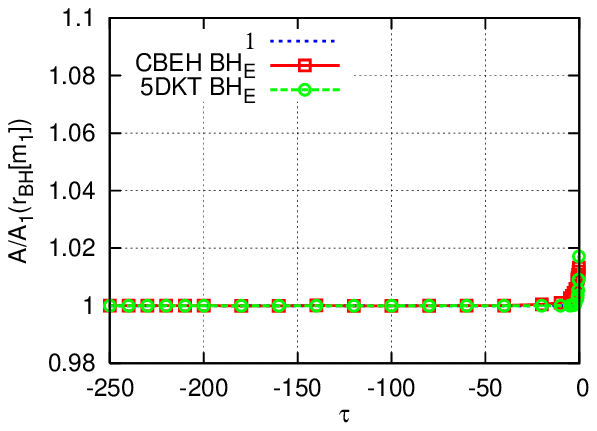}
\caption{Time evolution of area
of $\rm BH_E$s.
The vertical axis denotes the area normalized by ${\cal A}_1(r_{\rm BH}[m_1])$.
}
\label{fig:early-BH}
\end{center}
\end{figure}

\begin{figure}[htbp]
\begin{center}
\includegraphics[scale=1.5]{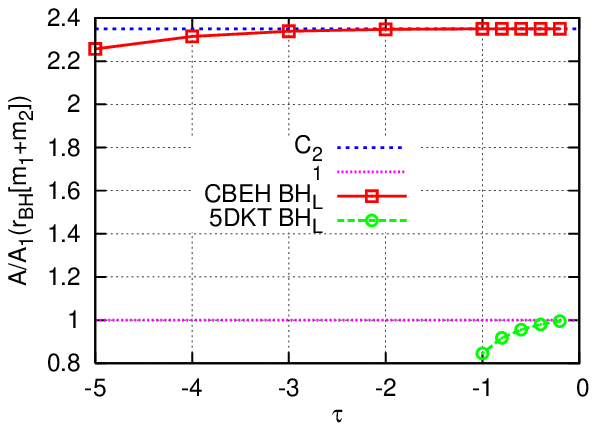}
\caption{Time evolution of the areas 
of ${\rm BH_L}$s.
The vertical axis denotes the area normalized by ${\cal A}_1(r_{\rm BH}[m_1+m_2])$.
}
\label{fig:late-BH}
\end{center}
\end{figure}

\begin{figure}[htbp]
\begin{center}
\includegraphics[scale=1.5]{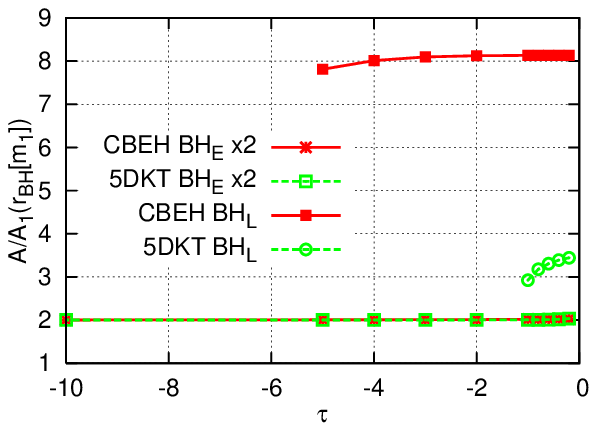}
\caption{Time evolution of area
of ${\rm BH_E}$ and ${\rm BH_L}$.
The vertical axis denotes the area normalized by ${\cal A}_1(r_{\rm BH}[m_1])$.
}
\label{fig:tot-BH}
\end{center}
\end{figure}

FIG. \ref{fig:early-dS} and FIG. \ref{fig:late-dS} 
show the evolution of the areas of de Sitter horizons 
$\rm dS_E$ and $\rm dS_L$. 
FIG.\ref{fig:early-BH} and FIG. \ref{fig:late-BH} show the time evolution of the areas of black hole horizons
$\rm BH_E$ and $\rm BH_L$. 
The time evolution of the areas of $\rm BH_E$ and $\rm BH_L$ is 
shown in FIG.\ref{fig:tot-BH}. 
In fact, from these figures, 
we can confirm that these values of areas asymptotically 
approach to the values computed from Eqs. (\ref{eq:rateE}) and (\ref{eq:rateL}). 
From FIG.\ref{fig:tot-BH}, we see that the area of ${\rm BH_L}$ in the CBEH 
at the appearance is larger than that in the 5DKT. 
It suggests that the impact parameter at the appearance of black holes in a spacetime with asymptotically locally Euclidean timeslices may be larger than that in a spacetime with asymptotically Euclidean timeslices.

\section{Summary and Discussion}\label{sec:5}
We have studied the evolution of marginal surfaces in 
the CBEH and the 5DKT. 
We have numerically searched for marginal surfaces in 
each time slice and calculated the areas
of the horizons. 
Each marginal surface corresponding to 
the black hole or de Sitter horizon at the early or the late time 
appears or disappears 
in pairs with another marginal surface. 
We have shown the time evolution of the marginal surfaces 
in Figs \ref{fig:tabkt}, \ref{fig:kthori1}, \ref{fig:kthori2}, \ref{fig:tabeh}
, \ref{fig:tabeh}, \ref{fig:ehhori1} and \ref{fig:ehhori2}.

The area at the appearance of the black hole enclosing 
both preexistent black holes 
in the CBEH is larger than that in the 5DKT. 
This suggests that the black hole production on the Eguchi-Hanson base space 
will be easier than that on the flat base space. 
It comes from the difference in the asymptotic structure between these solutions.
The results of this article give us the suggestion that the black hole dynamics may be notably affected by the topological structure of the extra-dimensions. 
In the context of TeV gravity scenarios, the topology of the bulk space might be nontrivial. 
Hence if our living higher dimensional world admits the asymptotic structure of
 the lens space topology, the black hole production rate in the linear collider might give us some information or the restriction to the model about the asymptotic structure of the extra-dimensions.

Although throughout this article, 
we focus on the time evolution of marginal surfaces on a certain timeslice, 
finally, we also comment on the event horizon. 
In fact, we have searched for the event horizon numerically 
by tracing null geodesics 
from the sufficiently future region to the past region~\cite{prep}. 
${\rm BH_E}$ and ${\rm BH_L}$ are almost 
identical to the cross sections of the event horizon 
with the timeslice $\tau=$constant surfaces at the sufficiently early and 
late time, respectively because the spacetime 
asymptotically becomes stationary in the sufficiently future and past regions. 

From a general viewpoint, 
Siino discussed the topology change of an event horizon 
in the four-dimensional spacetime~\cite{siino1,siino2} 
which is asymptotically stationary far 
in the future and showed that the non-trivial topology changes are 
caused by the set of endpoints of the event horizon, 
so-called, a crease set of the event horizon, 
where the event horizon is indifferentiable. 
Therefore, we expect that the difference between 
the topological structures of the crease sets will play 
a essential role in causing the difference in topology change 
in both solutions. In higher dimensional spacetimes, 
the structure of such crease sets is 
more complex than that in four-dimensions since the topology of an 
event horizon far in the future are not determined uniquely 
since the event horizons
in higher dimensional stationary spacetimes can admit 
various topologies~\cite{Cai,HelfgottGalloway}
in contrast to four-dimensional ones, which is restricted 
only to ${\rm S}^2$~\cite{Hawking}.
The detailed analysis about the event horizon  will be 
discussed in near future.

\section*{Acknowledgements}
We would like to thank K.Nakao for useful discussions.
C.Yoo is supported by the 21 COE program ``Constitution of wide-angle mathematical basis focused on knots" from Japan Ministry of Education. 
This work is supported by the Grant-in-Aid for Scientific Research No.13135208 and No.19540305.

\end{document}